\documentclass[aps,prb,showpacs,superscriptaddress,twocolumn]{revtex4-1}
\usepackage{graphicx}
\usepackage{amssymb}
\graphicspath{{images/}}

\begin{document}

\title{Anomalous transport in half-metallic ferromagnetic CrO$_2$}
\author{M. S. Anwar$^{1,2}$ and J. Aarts$^{1}$ \\${}^{1}$ Kamerlingh Onnes Laboratorium,
Leiden University, The Netherlands.\\${}^{2}$ Department of Physics, Kyoto
University, Kyoto 606-8502, Japan.}

\date{\today}

\begin{abstract}
We have investigated transport properties of CrO$_2$ thin films deposited on
TiO$_2$ and sapphire substrates. The films are good metals down to low
temperatures. The residual resistivity is of the order of 6 $\mu\Omega$cm for
films deposited on TiO$_2$ and two times higher for films on sapphire
substrates. The sign of the magnetoresistance (MR) changes from negative to
positive at a temperature around 100~K. This fact, as well as a rapid change in
the ordinary and anomalous Hall coefficients suggest a change in the electronic
state. At lower temperatures the MR is a linear function of the applied field.
This linear dependence might be explained as intergrain tunneling MR. This
interpretation is also suggested by the angular MR. The planar Hall effect
measurements reveal that the CrO$_2$ thin films are not in a single magnetic
domain state even for films deposited on an isostructural TiO$_2$ substrate.
\end{abstract}

\maketitle


\section{Introduction}

The material CrO$_2$ belongs to the class of half metallic ferromagnets (HMF)
\cite{goodenough71,schwarz86}, as revealed by electronic band structure
calculations \cite{lewis97,kroton98} and point contact Andreev Spectroscopy
(PCAS) \cite{soulen98,anguelouch01}; i.e. it has a gap in the minority-spin
density of states (DOS) ($N_{\downarrow}$) at the Fermi level of the order of
1.5~eV, but no gap in the majority DOS ($N_{\uparrow}$), resulting in complete
spin polarization at the Fermi level. These findings have stimulated the
interest in CrO$_{2}$ as a source of spin-polarized electrons for spintronics
devices. Also, its halfmetallic character was recently used for the realization
of long-ranged supercurrents \cite{keizer06,anwar10,anwar12}. However, the
electronic properties of CrO$_2$ are still not fully understood. For instance,
the resistivity between 10 K and 300 K is usually described in terms of an
excitation gap \cite{barry98,watts00}, but a clear connection with an
electronic or spin gap excitation cannot be made. Also, different results have
been reported with respect to the Hall effect. Watts {\it et al.} presented
data showing a sign reversal at low temperatures \cite{watts00}, which they
interpreted as evidence for two-band transport, but this was not found in later
studies \cite{yanagihara02,branford09}. In this article we return to the issue
of magnetotransport in high-quality thin films of CrO$_2$, with proper
attention to the different crystallographic axes of the material. We find
resistivity behavior which is subtly different from earlier reports, with an
anomaly around 100~K. We do not see a sign change in the Hall effect, but we do
find non-monotonous behavior in the high-field magnetoresistance. We also study
the low-field magnetoresistance behavior and come to a similar conclusion as
K\"{o}nig {\it et al.}, that Intergrain Tunneling Magnetoresistance (ITMR)
takes over from Anomalous Magnetoresistance (AMR) when the temperature
decreases to below 100~K. Data on the Planar Hall Effect (PHE) confirm that the
magnetization does not switch in single-domain fashion in these films,
different from one particular case reported by G\"{o}nnenwein
\cite{goennenwein07}. The article consists of two parts. First, the
measurements of the temperature-dependent resistance $R(T)$, of the high-field
MR and of the Hall effect are presented and discussed. Next, the data on the
low field magnetoresistance (MR) are given, with emphasis on the angular
dependent MR and on the planar Hall effect. We conclude that the data indicate
that a change in the electronic structure of CrO$_2$ takes place around 100~K,
possibly driven by a decrease of the carrier concentration.

\section{Material and sample preparation}
CrO$_2$ is a tetragonal material with a rutile structure and lattice parameters
$a$ = $b$ = 0.4421~nm and $c$ = 0.2916~nm. In CrO$_2$, the oxygen atoms form
octahedra around the Cr-atoms. There are two inequivalent octahedra,
side-sharing and corner-sharing ones. The side-sharing octahedra form a kind of
ribbons along the $c$-axis \cite{schlottmann03} (slightly distorted, elongation
along the $c$-axis~\cite{kroton98,lewis97}). The Cr ion in its formal 4+
valence state has two electrons in the $t_{2g}$ orbitals with the spin quantum
number $S=1$. As mentioned, CrO$_2$ is a HMF, although a Mott insulating-like
ground state and antiferromagnetic spin order could be expected because of
strong correlations. Kroton {\it et al.} \cite{kroton98} showed using the
LSDA+U method that the $d$ bands of CrO$_{2}$ are divided into two parts: a
weakly dispersing band well below the Fermi level and a strongly dispersing
band crossing the Fermi level. The former band provides the localized moments
and the latter is a strongly $s-d$ hybridized band that dilute the effect of
the $d-d$ Coulomb interaction and is responsible for the metallic behavior in
CrO$_{2}$. The oxygen $2p$ state extends to the Fermi level and plays the role
of electron or hole reservoirs. This causes self-doping and double-exchange
(DE) between the $d$-electrons and is responsible for the half-metallic nature.
A strong correlation between the spins of localized and non-localized electrons
makes the Hall effect and also the anomalous Hall effect a subtle tool to probe
topological spin defects of the 3D ferromagnetic material
 \cite{yanagihara02,branford09}. \\
The compound CrO2 is a metastable phase and bulk material is synthesized at
high pressures. Deposition techniques such as sputtering, pulsed laser
deposition or molecular beam epitaxy cannot be used, but high quality thin
films can be grown using the technique of chemical vapor deposition (CVD) at
ambient pressure, as for instance discussed in
Refs.~\cite{ishibashi79,li99,gupta99,rabe02,sousa06,miao06}. In CVD, a
precursor such as CrO$_{3}$ is thermally evaporated at 260~$^{\circ}$C and the
sublimated precursor transfers to a lattice matched substrate (such as
TiO$_{2}$ or Al$_{2}$O$_{3}$) at an elevated temperature of 390~$^{\circ}$C
using a pure Oxygen flow at 100~sccm. The lattice parameters of TiO$_2$ (rutile
with $a$ = $b$ = 0.4594 nm, mismatch with TiO$_2$ -3.8\%; $c$ = 0.2958~nm,
mismatch -1.5\%) closely match those of CrO$_2$ and epitaxial growth is
possible with small (although not negligible) effects of substrate-induced
strain. The growth on an $a$-axis oriented substrate is in the form of
rectangular grains with the long axis aligned along the film $c$-axis and the
short axis along the $b$-axis. It has been reported that pretreatment of the
TiO$_2$ substrates with hydrofluoric acid (HF) can enhance the strain in the
films \cite{miao06,anwar11,rameev04,rameev06} and it affects both magnetic and
electronic properties. Growth on sapphire is more complicated because of its
hexagonal structure ($a$ = 0.4754 nm), which is close to Cr$_2$O$_3$. Growth on
sapphire actually starts with Cr$_2$O$_3$ and then changes to the required
CrO$_2$ \cite{rabe02,sousa06}. Grains are aligned at 60$^\circ$ to each other
with six-fold rotational symmetry of hexagonal crystal structure of underlying
sapphire substrate. For details about film growth and morphology, see
Ref.~\cite{anwar11} \\
\begin{figure}[t]
        \begin{center}
\includegraphics[width=4cm]{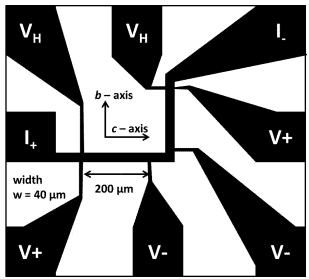}
\caption{Schematic of L-structure etched on CrO$_2$ thin films. Indicated are
the film crystal directions, the length and width of the bridges, and the
current and voltage contacts.} \label{sample}
    \end{center}
\end{figure}
To investigate transport properties of CrO$_2$ thin films, microbridges were
structured in the films deposited via the above mentioned CVD process on
untreated TiO$_2$, pretreated TiO$_2$ and sapphire substrates. For films
deposited on both pretreated and untreated TiO$_2$ substrates, L-shaped bridges
were fabricated in order to investigate the transport along both in-plane
crystal directions (current along the $b$- and $c$-axis) at the same time. They
were made with electron-beam lithography. The bridges were 40 $\mu$m wide, with
200~$\mu$m separation between the voltage contacts and 100~nm thickness of the
film. For the lithography step, a negative resist (MaN2405) was spin coated at
4000~rpm for 60~sec, and baked for 10~min at 90~$^{\circ}$C. Next, the
L-structure was etched in the CrO$_2$ films, a schematic is shown in
Fig.~\ref{sample}. It is difficult to etch the film with Ar ion etching because
of a rather slow etch rate. So, etching was done with reactive ion etching
(RIE), where a mixture of CF$_4$ (30~sccm) and O$_2$ (15~sccm) was utilized
with a background pressure of 10$^{-6}$~mbar. The RIE etch rate was of the
order of 0.8~nm/sec. For the sapphire substrate, a 200~nm thick film was grown,
in which a Hall bar (200~$\mu$m wide, 2~mm between the voltage contacts) was
structured with optical lithography.

\section{Results: Resistivity, magnetoresistance, Hall effect}

\subsection{Resistivity}

\begin{figure}[b]
        \begin{center}
\includegraphics[width=8cm]{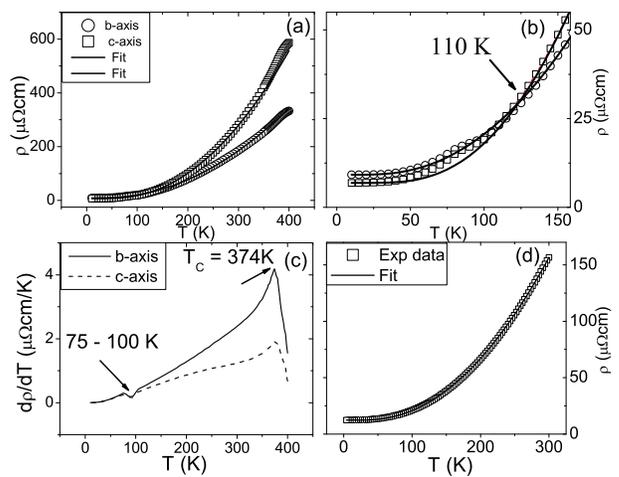}
    \caption{(a) Resistivity versus temperature for a 100~nm thick CrO$_{2}$
 film deposited on a pretreated TiO$_{2}$ substrate, along the in-plane
crystallographic $c$-axis (open squares) and the $b$-axis (open circles). The
solid lines are a fit of Eq. \ref{rt}, given in the text. (b) Crossover between
the resistivities at 110~K. (c) $d\rho/dT$ is showing the ferromagnetic
transition temperature at 374~K and a dip around 75-100~K along both $c$-
(dashed line) and $b$-axes (solid line). (d) Resistivity as a function of
temperature for a 200~nm thick CrO$_2$ film deposited on a sapphire substrate.
The solid line is the fit.}
    \label{RTtio2}
    \end{center}
\end{figure}

Figure \ref{RTtio2}a shows the resistivity as a function of temperature for a
100~nm thick CrO$_{2}$ film deposited on a pretreated TiO$_2$ substrate, along
both the $c$-axis and the $b$-axis. The residual resistivity ($\rho_{\circ}$)
is of the order of 9~$\mu\Omega$cm along the $b$-axis while along the $c$-axis
it is found 6~$\mu\Omega$cm. These values are quite similar to the literature
values ~\cite{watts00,yanagihara02}. It is noticeable that $\rho(T)$ at 4~K is
lower for the $c$-axis than for the $b$-axis, while this tendency reverses at
room temperature, with a crossover at 110~K (see Fig. \ref{RTtio2}b). We
observed an unexpected bump in $\rho(T)$ between 75 - 105~K along both in-plane
axes that is very clear in the derivative of the resistivity plotted in Fig.
\ref{RTtio2}c. The derivative also reveals the ferromagnetic transition at
around 374~K. Qualitatively, the results are the same for CrO$_{2}$ films
deposited on untreated TiO$_{2}$ and pretreated TiO$_{2}$ substrates although
there is rather a small quantitative difference.

Figure \ref{RTtio2}d presents $\rho(T)$ data of a 200~nm thick CrO$_{2}$ film
deposited on a sapphire substrate. At low temperature, $\rho(T)$ becomes almost
temperature independent, with $\rho_{\circ}$ $\approx$ 12~$\mu\Omega$cm, larger
than $\rho_{\circ}$ of the films deposited on TiO$_2$. In contrast, at room
temperature, $\rho$ is significantly lower than those for films on TiO$_2$.

In the literature, an accepted phenomenological expression used to describe
$\rho(T)$ is given by \cite{watts00,barry00},

\begin{eqnarray}\label{rt}
\rho(T)=\rho_{\circ}+AT^{2}e^{(-\frac{\Delta}{T})},
\end{eqnarray}

where $A$ is a coefficient. As shown in Fig. \ref{RTtio2} this expression fits
the $\rho(T)$ data well. Table \ref{resistivity-data} gives typical numbers for
$\rho_{\circ}$, $A$ and $\Delta$. The low values of $\rho_{\circ}$ indicate
that the films behave as good metals at low temperatures. The mean free path
$l_e$ can be estimated from the free electron model using the relation $l_e =
\frac{3}{e^2\rho_{\circ}\upsilon_{F}N}$, where $N$ is the density of states at
the Fermi level, $\upsilon_F$ is the Fermi velocity and $e$ is the charge of
the electron. Using $N = 7.55 \times 10^{46}$ states/J/cm$^3$ and $\upsilon_F =
2.5 \times 10^5$ m/s \cite{lewis97}, $l_e$ is evaluated to be about 100~nm.
This long $l_e$ suggests that the grain boundaries do not strongly affect the
transport behavior. The values of $\Delta$ are around 100~K, which does not
seem to be related to a characteristic energy scale of the material. This will
be discussed further below, but here we note that 100~K is the temperature
where $\rho(T)$ shows an anomaly.

\begin{figure}[h]
        \begin{center}
\includegraphics[width=8cm]{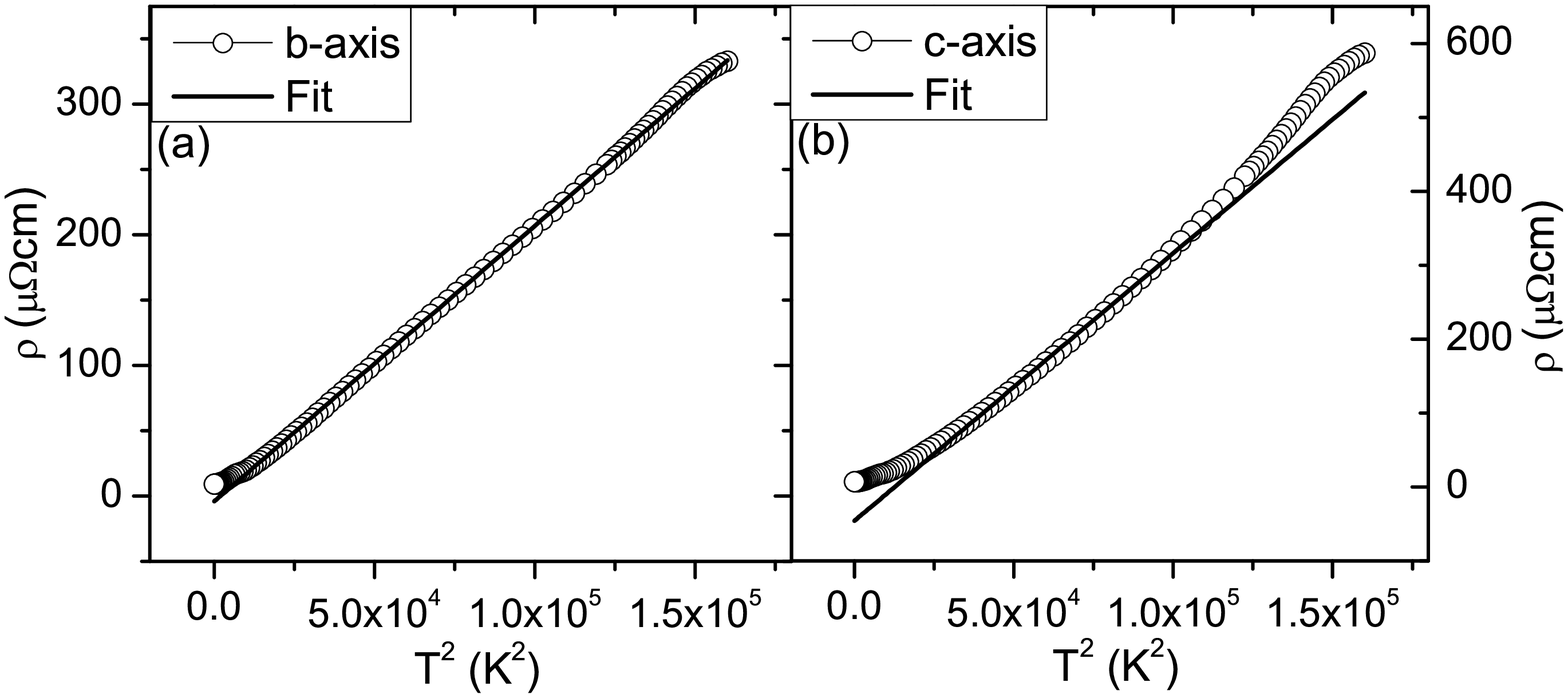}
    \caption{Resistivity versus $T^2$ for a 100~nm thick CrO$_2$ film deposited on
 TiO$_2$, (a) along the $b$-axis and (b) along the $c$-axis. The solid lines are
a fit of $\rho(T)=\rho_0+A^{'}T^2$ to the data.}
    \label{rtfits}
    \end{center}
\end{figure}

As the physical significance of $\Delta$ is not clear, we also tried to simply
fit a $T^{2}$ behavior $\rho(T)=\rho^{'}_{o}+A^{'}T^{2}$ without the
exponential term. The results are shown in Fig. \ref{rtfits}(a,b). For the
films deposited on TiO$_{2}$, resistivity shows a quite good fit to the T$^{2}$
dependence along the $b$-axis between 100~K to 350~K, similar to the results of
Suzuki {\it et al.} \cite{suzuki98} for CrO$_{2}$ film deposited on ZrO$_{2}$
substrate. In contrast, along the $c$-axis, the $T^{2}$ fit is quite poor and
is only successful between 215~K and 312~K. This fact is related to the change
in the anisotropic behavior of resistivity as a function of temperature.

\begin{table}[h!]
    \begin{center}
\caption{Some important parameters $\rho_{\circ}$, $\Delta$ and $A$ for CrO$_{2}$
thin films deposited on pretreated, untreated TiO$_2$ and sapphire substrates.}\label{resistivity-data}
\begin{tabular}{|c|c|c|c|}\hline
  Samples & $\rho_{\circ}$ & $\Delta$ & $A$ \\
      &$\mu\Omega$cm&  (K)& n$\Omega$cm /K$^{2}$\\\hline
    pretreated-TiO$_{2}$
        ($c$-axis)& 6 & 80 & 2.8 \\
        ($b$-axis) & 9 & 150 & 5.2\\\hline
    untreated-TiO$_{2}$ 
        ($c$-axis)& 7 & 75 & 2.6 \\
        ($b$-axis) & 11 & 140 & 3.9 \\\hline
       sapphire & 12 & 90 & 2.2 \\\hline
    \end{tabular}
\end{center}
\end{table}

\subsection{Magnetoresistance: High field MR}

Magnetoresistance (MR) is the measure of the relative change in the resistance
of a material in an externally applied magnetic field at a constant
temperature, and defined as
\begin{eqnarray}
    MR = \frac{\Delta\rho}{\rho} = \frac{R(H)-R(0)}{R(0)},
    \label{AMR}
\end{eqnarray}

where $R(0)$ is the resistance in zero field and $R(H)$ is the resistance in
the field. High-field MR was measured at different temperatures with a
commercial apparatus (Quantum design, PPMS) with fields (maximum $\pm$ 9~T)
oriented along the out-of-plane direction for CrO$_{2}$ films deposited on both
pretreated and untreated TiO$_{2}$ substrates.

\begin{figure}
        \begin{center}
\includegraphics[width=8cm]{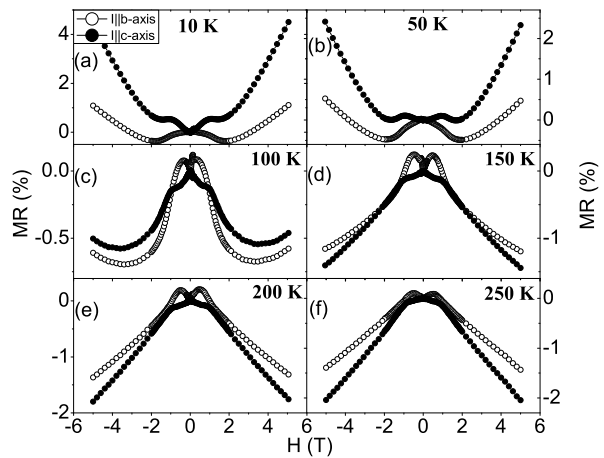}
    \caption{Magnetoresistance as a function of applied field for a 100~nm thick
    CrO$_{2}$ film deposited on a pretreated TiO$_{2}$ substrate, for various temperatures.
    The field is perpendicular to the substrate and the current $I$ is either along the
    $b$-(open circles) or the $c$-axis (close circles).}
\label{MR-HF-tio2}
    \end{center}
\end{figure}

\begin{figure}
        \begin{center}
\includegraphics[width=8cm]{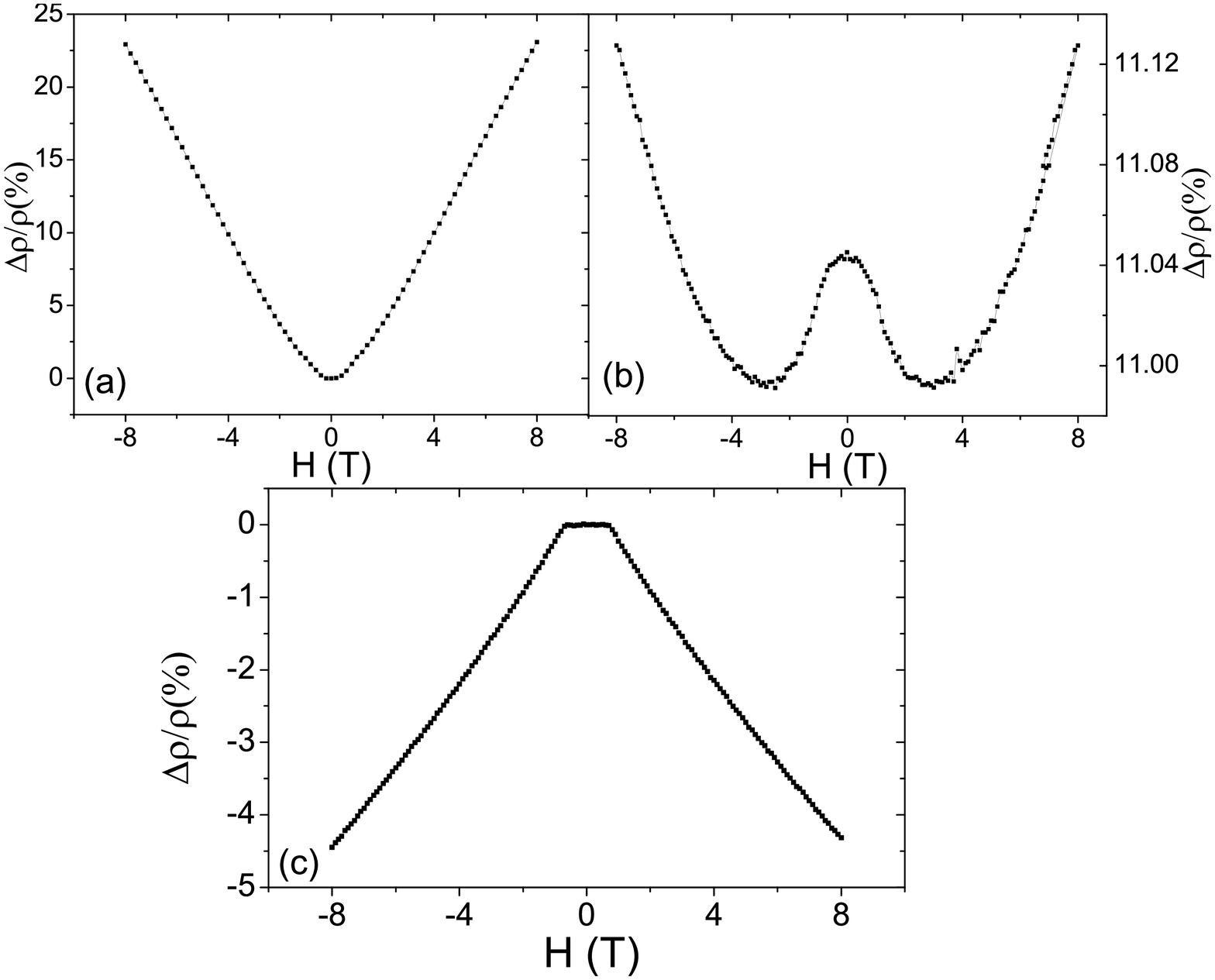}
    \caption{Magnetoresistance for a 200~nm thick CrO$_{2}$ thin film deposited on a
    sapphire substrate up to $\pm$9~T magnetic field at different temperatures,
    (a) 10~K, (b) 100~K and (c) 300~K.}\label{AMRHsapphire}
    \end{center}
\end{figure}

Figure~\ref{MR-HF-tio2} shows the data for MR measured at various temperatures
between 10~K to 250~K for a 100~nm thick CrO$_{2}$ film deposited on a
pretreated TiO$_2$ substrate. At low fields, the MR shows variations associated
with the changes in the magnetization. When the magnetization saturates, above
1~T, monotonic behavior sets in. An interesting observations is that the slope
of MR in high fields is different in sign for low temperatures and high
temperatures, and that at temperatures below 150~K the crossover can be seen as
function of field. Above 100~K, the MR is negative, with values around -2$\%$
at 5~T around room temperature (250~K). At 100~K, the sign is still negative
but a cross-over to positive behavior is visible at about 4~T, where MR changes
quadratically. At 50~K and 150~K, the MR starts to be negative, but reverses to
positive around 2~T. At 10~K the MR reaches 4$\%$ (1$\%$) in 5~T with current
along the $c$-axis ($b$-axis).We observed similar behavior for films on
untreated TiO$_{2}$, except that the lower field curves are not symmetric. This
symmetry might be related to the quality of the films.

Figure \ref{AMRHsapphire} shows the MR for the 200~nm thick film on sapphire,
again with the field applied to the out-of-plane direction. The data show the
same features; positive MR at 10~K, a cross-over at 100~K, negative MR at
300~K. Noteworthy are the large values at 10~K, of the order of 30$\%$ at 8~T.

\subsection{Anomalous Hall Effect}

\begin{figure}
        \begin{center}
\includegraphics[width=8cm]{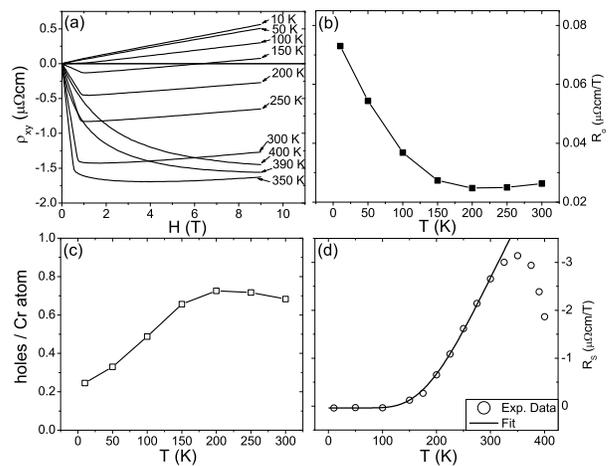}
    \caption{(a) Hall resistivity versus applied field $\rho_{xy}(H)$ for a 100~nm
    thick CrO$_{2}$ film deposited on a pretreated TiO$_{2}$ substrate, measured at
    various temperatures between 10~K to 400~K. (b) Ordinary Hall coefficient $R_{\circ}$
    as a function of temperature between 10 - 300~K. (c) Number of holes/Cr atom versus
    temperature decreases with the increase in the temperature. (d) Anomalous Hall
    coefficient $R_S$ versus temperature, the solid line is the theoretical fit
    using Eq. (\ref{Rs}).}\label{AHE}
    \end{center}
\end{figure}

Figure \ref{AHE}a shows the Hall resistivity $\rho_{xy} = V_yw/I_x$ (where $w$
is the width of the Hall bar) as a function of externally applied field in
out-of-plane configuration, for various temperatures between 10~K to 400~K,
using the same L structure. The measurement was done for films deposited on
both pretreated and untreated TiO$_2$ substrates, and for current passing along
both the $c$- and the $b$-axis. We did not observe any difference beyond the
experimental error for both directions of current and for both kind of films,
in agreement with Onsager's principle that $\rho_{xy}=\rho_{yx}$ regardless of
crystal orientation.

At low temperatures ($<$50 K), $\rho_{xy}(H)$ is linear with a slope that
corresponds to hole-like charge carriers. Between 100~K and 350~K, an extra
contribution is visible at low fields, which is usually ascribed to the effects
of the magnetization, and referred to as the anomalous Hall effect (AHE).

The Hall resistivity can then be written as

\begin{eqnarray}
\rho_{xy} = \mu_{0}(R_{0}H_a + R_{S}M) \label{ahe}
\end{eqnarray}

with $R_{0}$ the normal Hall coefficient and $R_S$ the anomalous Hall
coefficient.The carrier density $n$ follows in a one-band model from $R_{0} =
-1/en$. A positive $R_0$ corresponds to holes as carriers.

\section{Discussion}
The discussion on the data given above can be started with $\rho_{\circ}$. The
different values of $\rho_{\circ}$ for different substrates indicates a
substrate dependence of the film quality because $\rho_{\circ}$ is sensitive to
the disorder. The ratio between room temperature resistivity and $\rho_{\circ}$
is a measure of the crystal imperfections or impurity concentration as
electron-phonon scattering vanishes at low temperatures. This ratio is known as
the residual resistance ratio (RRR) \cite{kittel96}. For our samples the RRR is
20 along the $b$-axis and 41 along the $c$-axis. These values are higher than
those for the films deposited on an untreated TiO$_2$ and sapphire substrates.
This fact reveals that CrO$_{2}$ films deposited on a pretreated TiO$_2$
substrate are of better quality.

Another important issue is the description of $R(T)$ with Eq. (\ref{rt}), which
is usually interpreted as a $T^2$ contribution modified with a phenomenological
exponential. In general, the $T^2$ term is attributed to electron-electron
scattering. The value of the coefficient $A$ of the $T^{2}$ term is in the
range of 2.2 - 5.0 $\times$ 10$^{-3}$ $\mu\Omega$cm/K$^{2}$ and much larger
than those for ordinary ferromagnetic metals (e.g. 1.3-1.6 $\times$
10$^{-5}$~$\mu$$\Omega$cm/K$^{2}$; for Ni, Fe)~\cite{suzuki98}. The higher
value might be related to the contribution of the electron-magnon scattering
along with electron-electron scattering~\cite{barry00,suzuki98}. If $\rho(T)$
also has electron-magnon scattering contributions then the prefactor $\Delta$
of the exponential term might be related with a gap in magnon spectrum.
However, the value of $\Delta$ is found to be about $\approx$ 150~K (maximum,
along the $c$-axis), which is still too low to be associated with spin flip
scattering, since the minority spin band is about 1.5~eV below the Fermi level.
That suggests there is no correlation of $\Delta$ with spin flip scattering in
CrO$_2$.
It is remarkable that the value of $\Delta$ falls in the temperature range of
about 100~K where we find a dip in $dR/dT$. This suggests a certain electronic
phase change in CrO$_2$ around 100~K. That is reinforced by the high-field MR
data, which show a field dependent sign change around 100~K which was not
observed in earlier work \cite{watts00}. In contrast, we do not find a sign
reversal in the Hall data, which allows us to extract a carrier density using a
one-band model as shown in Fig.~\ref{AHE}(c). The carriers are holes, and we
find that their number actually is not constant: $n$ starts to drop
significantly for temperatures below 200 K. There appears therefore to be no
reason to try describing the resistivity in the whole temperature regime with a
single expression. Something can also be said about the anomalous Hall
coefficient $R_S$, which is plotted in Fig. \ref{AHE}(d). Although $R_S$ is
negligibly small below 100~K, it grows exponentially around 150~K and has a
peak at 350~K, just below the Curie temperature. It is also interesting that
the sign of $R_{0}$ and $R_S$ are different, since for conventional
ferromagnets the signs are the same. The different sign is quite similar to
what has been observed in Colossal Magnetoresistance materials such as
(La$_{0.7}$Sr$_{0.3}$)MnO$_3$ or (La$_{0.7}$Ca$_{0.3}$)MnO$_3$ and they seem to
rule out the conventional explanations of conventional spin scattering and side
jump or skew scattering but a Berry phase might be the possible explanations
for these materials. Recently, it was suggested that topological spin defects
or Skyrmion strings \cite{yanagihara02,ye99} can be an origin of the behavior
of AHE, in particular for double exchange systems (also the case of CrO$_2$
with self doped double exchange). The density of Skyrmion strings $n^{*}$ and
$R_S$ are related as

\begin{eqnarray}
R_{S}\propto \frac{1}{T}<n^{*}> \propto
\frac{\texttt{exp}(E_{C}/k_{B}T)}{T}\label{Rs}
\end{eqnarray}

where $E_C$ is the energy for creating a single Skyrmion string. In our data,
$R_S$ increases exponentially around 150~K and yields a good agreement with Eq.
(\ref{Rs}) with $E_C$ $\approx$ 1100~K (see Fig. \ref{AHE}(d)). Concluding this
section, we come to a somewhat different picture for the electronic structure
of CrO$_2$

\section{Results: Low field MR, Rotational scans of MR, Planar Hall Effect}

\subsection{Magnetoresistance: Low field MR}

The low field MR was measured at 4.2~K with a cryostat (Oxford instruments
$\mu$ metal shielded) with externally applied magnetic field with in-plane
configuration. For the same samples used in above mentioned experiments, we
applied field parallel and perpendicular to the current for both cases of
current along the $c$- and the $b$-axis. The field $H$ was applied parallel to
the current $I$ for the film deposited on sapphire with the Hall bar structure.
For all cases, four probe dc measurements with a current of 100~$\mu$A were
used.

\begin{figure}
        \begin{center}
\includegraphics[width=8cm]{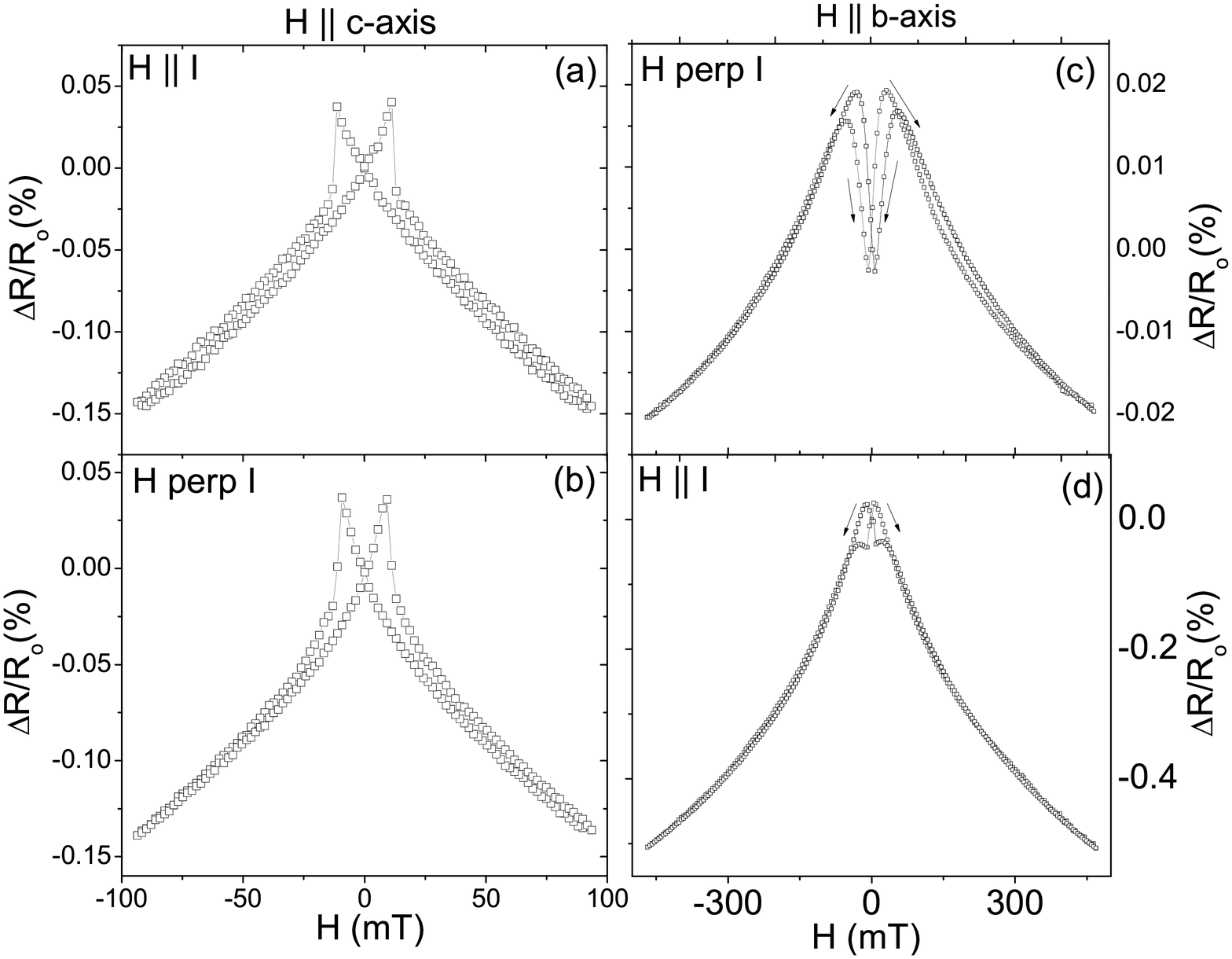}
    \caption{Low field MR probed at 4.2~K on a 100~nm thick CrO$_{2}$ film deposited
    on a pretreated TiO$_2$ and simultaneously measured for both cases of $I\| c$-axis
    and $I \| b$-axis. (a) $H \| I \| c$-axis, (b) $H \bot I \| c$-axis ($H$ along the
    $b$-axis), (c) $H \bot I \| b$-axis ($H \| c$-axis) and (d) $H \| I \| b$-axis.}
    \label{RHtreated}
    \end{center}
\end{figure}

In all cases, the resistance increases when coming from high field, and shows a
hysteretic structure when the magnetization direction switches and domain
forms. When field is applied along the $c$-axis (H$ \| c$) then for both H$ \|
I$ (or $I \|c$) and H$ \bot I$ (or $I \| b$) the  data show a jump-like
decrease of $R$ at the presumed coercive field $H_c$ (see Fig.
\ref{RHtreated}(a)(b)). When the field is applied along the $b$-axis (H$ \| b$)
then the resistance for H$ \bot I$ (or $I \| c$) exhibits a dip slightly above
$H_c$ and a peak around $H_c$ (see Fig. \ref{RHtreated}c). For H$ \| I$ (or $I
\| b$) a different structure is seen with a plateau slightly above $H_c$ and a
peak at $H_c$(see Fig. \ref{RHtreated}(d)).

The MR behavior was already studied by K\"{o}nig {\it et al.} \cite{koenig07},
with results similar to these. They interpreted their results assuming the
$c$-axis as easy axis; regardless of the angle between $H$ and $I$, the
magnetization switches sharply for H$ \| c$. For H$ \| b$ domains start to form
well above $H_c$, which leads to a dip or a plateau in the variation of $R$.
For their sample, a magnetization measurement confirmed that the $c$-axis is
indeed the easy axis.

\begin{figure}
        \begin{center}
\includegraphics[width=8cm]{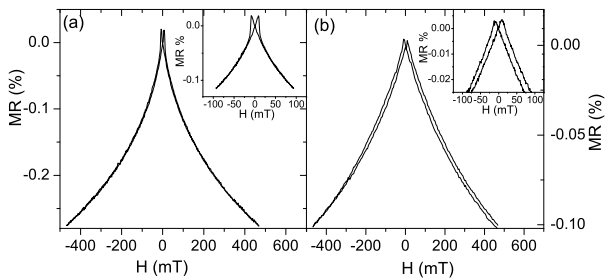}
    \caption{Low field and low temperature (4.2~K) MR for a 200~nm thick CrO$_{2}$
    thin film deposited on a sapphire substrate, (a) $H \| I$ (b) $H \bot I$. The
    insets show the MR for field up to 100~mT.}
    \label{AMR-sapphire}
    \end{center}
\end{figure}

Figure \ref{AMR-sapphire}(a) presents the low field MR data for a 200~nm thick
CrO$_2$ film deposited on a sapphire substrate for $H \| I$. The MR is negative
with a sublinear decrease up to 0.25$\%$, which is similar to the MR data of
CrO$_2$ films deposited on TiO$_2$ substrates for $H \bot I$. The AMR peaks
around the coercive field are obviously present (see the inset of Fig.
\ref{AMR-sapphire}(a)). The MR for the perpendicular configuration is two times
less than the MR for the parallel configuration of applied field. The peaks at
the coercive field are also very weak for $H \bot I$ (see Fig.
\ref{AMR-sapphire}(b)) but the decrease is still sublinear.

\subsection{Rotational scans of MR}

\begin{figure}
        \begin{center}
\includegraphics[width=8cm]{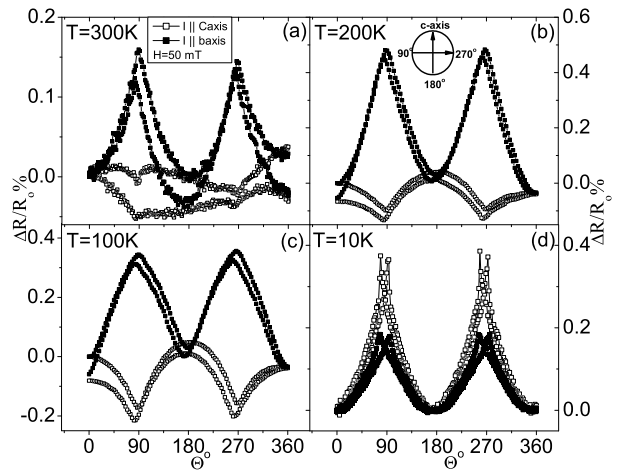}
    \caption{Relative change in the resistivity as a function of rotation of
    applied field of 50~mT. The data are taken for CrO$_2$ film on a pretreated TiO$_2$
    substrate (a) at 300~K (b) 200~K, (c) 100~K and (d) at 10~K. We define $\theta$ as
    the angle between the magnetic field and the $c$-axis as shown in the inset of (b).}
    \label{rotationscan}
    \end{center}
\end{figure}

We also measured the MR as a function of the angle $\theta$ of the applied
field with respect the $c$-axis for $H || b$ and $H || c$. We probed
$R(\theta)$ at different temperatures and also at various magnetic field
strengths using a rotational sample holder of PPMS Quantum design. In
Fig.~\ref{rotationscan}, $R(\theta)$ at different temperatures for 50~mT
applied field is plotted. The data for both configurations of $I \| c$ and $I
\| b$ are simultaneously recorded. At $\theta = 0$ the applied field is along
the $c$-axis as shown in the inset of Fig.~\ref{rotationscan}(b).

\begin{figure}
        \begin{center}
\includegraphics[width=8cm]{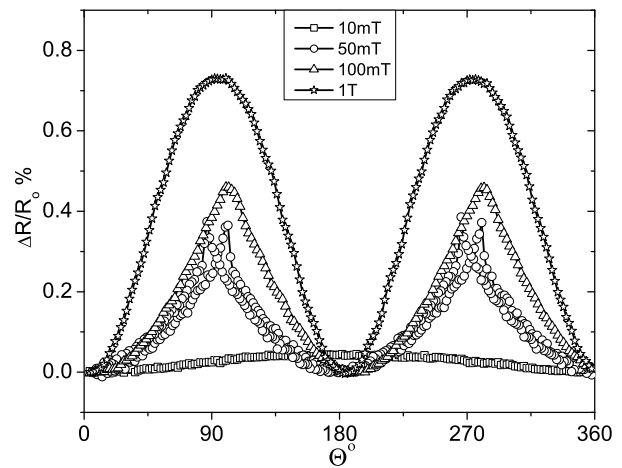}
    \caption{Rotational scans of magnetoresistance of a CrO$_2$ film on a pretreated
    TiO$_2$ substrate at various fields at 4.2~K for current $I \| c$-axis.}
    \label{R(oh)}
    \end{center}
\end{figure}

At 300~K, $R(\theta)$ for $I \| c$ is weakly varying, with signatures of maxima
at 0$^{\circ}$, 180$^{\circ}$, and minima at 90$^{\circ}$, 270$^{\circ}$. For
$I \| b$ there is a clear variation with peak-like maxima at 90$^{\circ}$ and
270$^{\circ}$, and rounded minima at 0$^{\circ}$ and 180$^{\circ}$. At 200~K
the data are similar, now with a stronger variation for $I \| c$. At 100~K the
peaks become round somewhat, but there is no qualitative change. At 10~K,
although, the data for $I \| b$ are still similar, the data for $I \| c$
exhibit strong difference: the minima at 90$^{\circ}$ and 270$^{\circ}$ have
converted to sharply peaked maxima, similar to the $I \| b$ data. The shape of
the maxima, and the small hysteresis which can be seen to develop, are partly
due to the relatively small applied field. For larger fields the MR-effect
becomes stronger, and the maxima more rounded, as shown in Fig. \ref{R(oh)} for
$I \| c$ at 4.2~K.

Also these observations are similar to earlier ones \cite{koenig07}. To
understand what happens, we compare the 100~K data with the 10~K data. At 100~K
the behavior can be explained with the $c$-axis being the easy axis. It yields
a maximum at 0$^{\circ}$ for H$ \| c \| I$, since the magnetization is parallel
to the current, which gives a higher $R$. Also at 90$^{\circ}$, the
configuration H$ \| c \| I$ gives domains with a magnetization perpendicular to
the current, and therefore a minimum in $R$. At 10~K the effect of the easy
axis seems to have disappeared and the parallel alignment of magnetization and
current (the situation H$ \| c \| I$) now leads to \emph{minimum}. This can be
explained by assuming that the dominating transport mechanism is ITMR. The
parallel alignment of the magnetization of neighboring grains \emph{reduces}
the scattering at grain boundaries. It is obvious that this effect can be
particularly relevant for fully spin-polarized materials. It also shows a
definite influence of the grain boundaries in our thin films on the electrical
transport properties.

\subsection{Planar Hall effect}

\begin{figure}
        \begin{center}
\includegraphics[width=8cm]{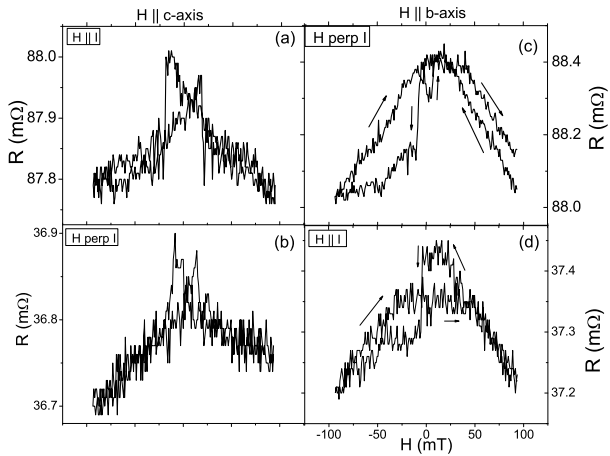}
    \caption{Planar Hall effect for a 100~nm thick CrO$_{2}$ film deposited on a
    pretreated TiO$_{2}$ substrate, (a-b) $H \| c$-axis (c-d) $H \| b$-axis, at 4.2~K.}
    \label{PHE}
    \end{center}
\end{figure}

The resistance measured along the direction of the current as a function of
applied field is known as anisotropic magnetoresistance (AMR), but this
physical mechanism is also responsible for a Hall voltage, or Hall resistance,
i.e. in the direction perpendicular to the applied current and field. This Hall
voltage is commonly called Planar Hall Effect (PHE). The only report in the
literature on PHE measurements for CrO$_2$ films showed that at intermediate
thickness (100~nm), films can develop biaxial magnetic anisotropy, in which two
magnetic easy axes occur, one in between the $c$-axis and the $b$-axis, and one
mirrored around the $c$-axis to lie in between the $c$-axis and the $b$-axis
\cite{goennenwein07}. Moreover, they also predict that their films are in a
single magnetic domain structure. We also probed PHE using the L structure of a
100~nm thick film at 4.2~K in the shielded cryostat with a magnetic field
applied in a parallel configuration (H$ \| I$) but our film was exhibiting
uniaxial magnetic anisotropy, for details see Ref. \cite{anwar11}. The
transverse voltages were recorded for $I || c$ and $I || b$ when the H$ || c$
and for the H$ \| b$. The results are given in Fig. \ref{PHE} for all four
different configurations of current and field.

Comparing Fig. \ref{RHtreated} with Fig. \ref{PHE} we see that the PHE signal
is strongly correlated with the AMR behavior for H$ \| c$ (the easy axis of
magnetization). Both show narrow peaks, e.g. switching behavior, at the
coercive field $H_c$. For H$ \| b$ there is less resemblance with AMR. There is
no dip-peak structure for H $\bot I$; for H $\| I$ there is a weak signature of
plateau-peak.For films deposited on untreated substrates we did not observe any
PHE signal. This fact suggests that the PHE is quite sensitive to disorder.

\begin{figure}
        \begin{center}
\includegraphics[width=8cm]{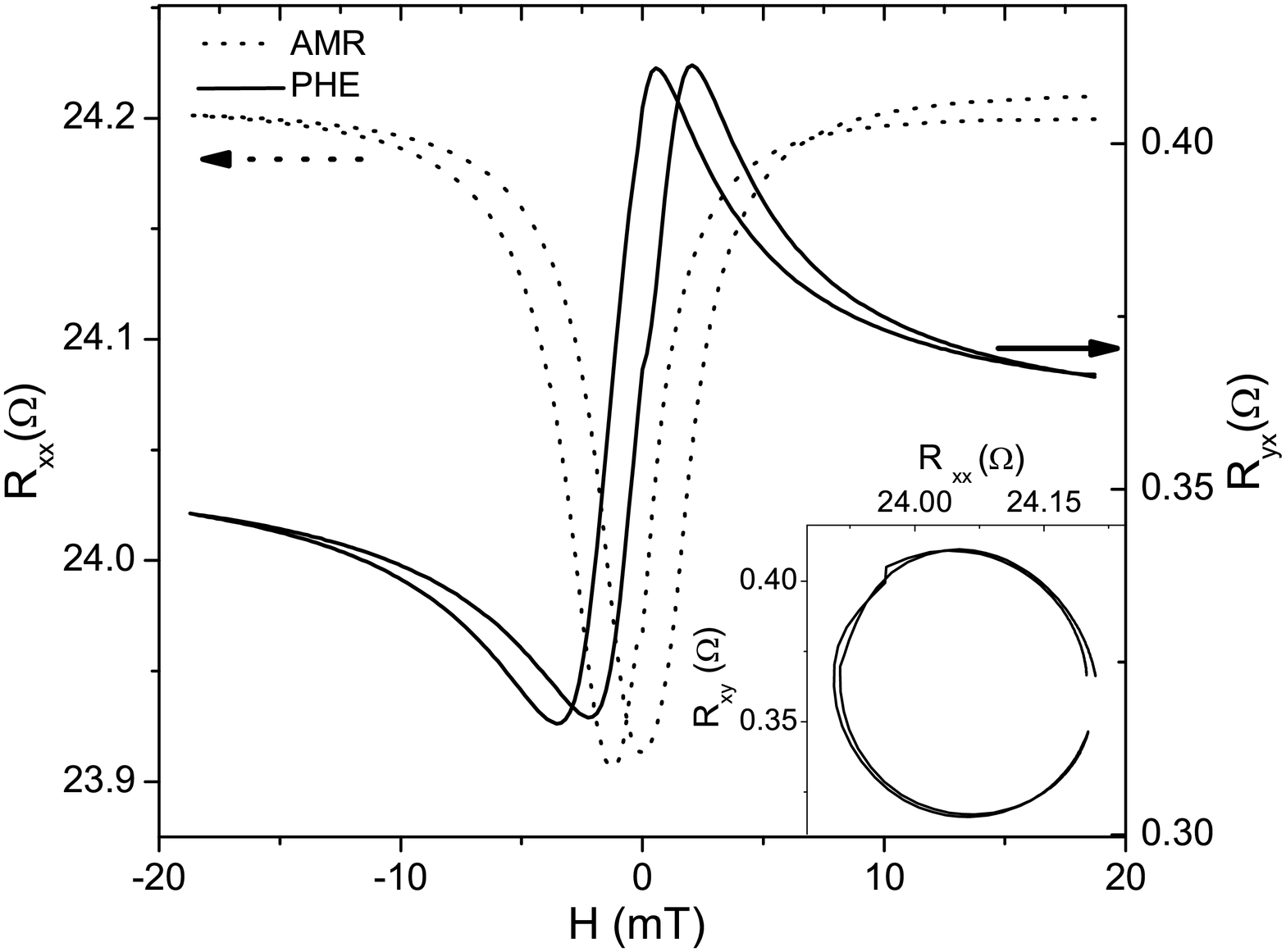}
    \caption{Planar Hall effect ($\rho_{xy}$) and AMR ($\rho_{xx}$) at 4.2~K for a
    20~nm thick Permalloy thin film deposited on a Si substrate. The inset shows the
    correlation between the $\rho_{xx}$ and $\rho_{xy}$, the circle formation shows
    the single domain structure for Py film.}
    \label{PHE-py}
    \end{center}
\end{figure}

The interest in PHE stems from the fact that, if magnetic structures are in a
single domain, the longitudinal electric field $E_x$ (measured by AMR) and the
transverse field $E_y$ (from PHE) are given by

\begin{eqnarray}
E_x&=&\bigg[\frac{\rho_\parallel+\rho_\perp}{2} + \frac{\rho_\parallel-\rho_\perp}{2} \cos 2\theta \bigg]J,\\
E_y &=&\bigg[ \frac{\rho_\parallel-\rho_\perp}{2} \sin 2\theta\bigg]J,
\label{AMReq}
\end{eqnarray}

where $\theta$ is the magnetization angle. Plotting $E_y$ against $E_x$, the
resulting graph should be a circle if the magnetization rotates as a single
domain. The magnetization angle can then be extracted for every value of
($E_x,E_y$). An example is illustrated in Fig. \ref{PHE-py} for a 20~nm thick
permalloy (Py) film measured at room temperature. Looking at CrO$_2$, it is
obvious that the plot of ($E_x,E_y$) will not form a circle. This might
indicate that the material is not in a single domain state at the measured
temperature of 4~K. In the view of the rotational scans, it seems more logical
to conclude that the PHE data confirm the conclusion that the low temperature
magnetotransport is dominated by ITMR and not by AMR.

\section{Conclusion}

It can be concluded that CrO$_2$ thin films deposited on a pretreated
TiO$_2$substrate are of better quality than the films deposited on an untreated
TiO$_2$ or on sapphire substrate. The higher value of the coefficient of the
$T^2$ dependence of the resistivity might come from electron-magnon scattering
along with the electron-electron scattering. The bump in $R(T)$ and the sign
change in MR around 100~K appear to be related with some change in the
electronic configuration of CrO$_2$, possibly driven by the decrease in carrier
concentrations as found in the Hall data. The phenomenological quantity
$\Delta$ used to describe $R(T)$ might also be connected to this change in
electronic structure and carrier concentration, rather than with the magnon gap
or spin flip scattering. The low-field MR and PHE data reveal the presence of
ITMR, and stress the presence of grain boundaries in our films. It is
remarkable that the change in electronic structure and the change from AMR to
ITMR take place in roughly the same temperature region, but since the size of
the grains is much larger than the typical mean free paths, it would appear
that the grain boundaries cannot have a decisive influence on the electronic
behavior, and both phenomena are unrelated.

\section*{Acknowledgement}
We are grateful to Shingo Yonezawa for fruitful discussions. M. S. A. is
thankful to the Higher Education Commission Pakistan for financial support.
This work was part of the research program of the "Stichting voor Fundamenteel
Onderzoek der Materie (FOM)" which is financially supported by the "Nederlandse
Organisatie voor Wetenschappelijk Onderzoek (NWO)."


\end{document}